\documentclass[11pt,twocolumn]{article}

\usepackage[utf8]{inputenc} 
\usepackage[T1]{fontenc}    
\usepackage{hyperref}       
\usepackage{url}            
\usepackage{booktabs}       
\usepackage{amsfonts}       
\usepackage{nicefrac}       
\usepackage{microtype}      
\usepackage{lipsum}
\usepackage{amsmath}
\usepackage{amsfonts}
\usepackage{amssymb}
\usepackage{makeidx}
\usepackage{graphicx}
\usepackage{lmodern}
\usepackage{kpfonts}
\usepackage{pdfpages}
\usepackage{lscape}
\usepackage{multicol}
\usepackage{rotating}
\usepackage{comment}
\usepackage{float}
\usepackage{authblk}
\usepackage{setspace}

\newcommand{\PH}[1]{{\color{black} #1}}
\newcommand{\AB}[1]{{\color{black} #1}}

\providecommand{\keywords}[1]
{
  \small	
  \textbf{\textit{Keywords---}} #1
}

\title{
Modelling of flow through spatially varying porous media with application to topology optimization
}

\author[1]{Michaël Rakotobe\thanks{Corresponding author: \texttt{michael.rakotobe@univ-reunion.fr}}}
\author[1]{Delphine Ramalingom}
\author[1]{Pierre-Henri Cocquet}
\author[1]{Alain Bastide}
 
\affil[1]{Université de La Réunion, Laboratoire PIMENT, 117 rue du Général Ailleret, Le Tampon 97430, France}

\date{}

\begin{document}
\maketitle
\begin{abstract}
The objective of this study is to highlight the effect of porosity variation in a topology optimization process in the field of fluid dynamics. Usually a penalization term added to momentum equation provides to get material distribution. Every time material is added inside the computational domain, there is creation of new fluid-solid interfaces and apparition of gradient of porosity. \PH{However,} at present, porosity variation is not taken account in topology optimization \PH{and} the penalization term \PH{used to locate the solid} is analogous to a Darcy term used for flows in porous media. With that in mind, \PH{in this paper,} we first develop an original one-domain macroscopic model for the \PH{modelling} of flow through \PH{spatially} varying porous media \PH{that goes} beyond the scope of Darcy regime. \PH{Next, we numerically solve a topology optimization problem and compare the results obtained with the standard model that does not include effect of porosity variation with those obtained with our model.}
Among our results, we show for instance that the designs obtained are different but percentages of reduction of objective functional remain quite close (below 4\% of difference). In addition, we illustrate effects of porosity and particle diameter values on final optimized designs.
\end{abstract}


\section{Introduction}

\PH{
Topology optimization usually aims at finding the location of a solid inside a fluid that either maximize or minimize a given cost functional. 
The solid zones are located thanks to a penalization term added to momentum equation which vanishes in the fluid zones and goes to infinity in the solid zone. This term is analogous to a Darcy term used for flows in porous media. While such problems applied to fluid flow originally concerned Stokes flows \cite{borrvall2003topology}, it has then been applied to several other setting like for instance heat transfer in fluid flow \cite{qian2016topology,subramaniam2019topology}, turbulent flows \cite{papoutsis2016continuous,yoon2016topology} or buoyancy-driven flows
\cite{alexandersen2014topology,ramalingom2019multi}.
We also would like to refer to the two following review papers \cite{dbouk2017review,Alexandersen2020review} for many other references on this topic. 

In this paper, we wish to perform topology optimization using a porous media that only slow down the flow instead of blocking it. Several references dealing with 
such problems can be found in \cite[p. 13, Section 2.5]{Alexandersen2020review}.
More precisely one can find, in \cite{bastide2018penalization}, a penalization model allowing to get optimized porous media modelled with Darcy law.  
In \cite{shen2018topology}, topology optimization problem with either fluid/porous or solid phases is investigated. They used the penalized Navier-Stokes model with a penalization parameter that depends on the Darcy number $\mathrm{Da}$ hence fluid corresponds to large $\mathrm{Da}$, solid to small $\mathrm{Da}$ and porous media to intermediate values. In \cite{philippi2015topology,takezawa2019method}, 
the inertial/nonlinear effect of the flow in the porous media are taken into account thanks to the Darcy-Brinkman-Forchheimer model and the porosity of the medium is then used as optimization parameter. We emphasize that looking for optimized porous medium is going to produce designs made by some isolated pieces of porous media inside the fluid, we then have to deal with the variation (gradient) of porosity in our mathematical model. However, to the best of our knowlege and in the aformentioned references, this feature has not been taken into account.} 

The modelling of flow through porous media is usually done with the volume averaging method presented in \cite{Whitaker1999}, where the concept of a representative elementary volume (REV) is introduced. However, we can see \cite{Nield2017} that there are situations where the notion of volume averaging is no longer applicable due to effects of porosity variation. It is the case for example in the so-called channeling effect where in the vicinity of the wall, porosity is near unity before reaching nearly its core value at about five diameters from the wall \cite{Nield2017}. A solution might be to consider a deforming averaging volume and to deal with porosity gradients which are explicitly present within the macroscopic momentum equation as explained in \cite{Goyeau1997}. In this study, we adopt this point of view and although complex local closure problem were pointed out by \cite{Goyeau1997}, we did not have this issue thanks to assumptions taken from the outset in following the development in \cite{Diersch2014}.
\newline
Another difficulty encounter in the modelling of porous media is the numerical treatment of interfaces between fluid and porous domains. Many authors introduced special treatment of pressure and velocity at the interface \cite{Betchen2006} \cite{DeGroot2011} that are physically reasoned to avoid unphysical numerical oscillations. In this study, we use a smooth transition of the porosity between fluid and porous domains in considering the increase of packing density with depth. This physical conception is supported by the findings of \cite{Khalili2014} where the impact of depth-dependent porosity was experimentally and numerically studied and a porosity-depth relation has been obtained.

\PH{
Owning to the previous literature review, this paper is going to be dedicated to study the influence of porosity variation in topology optimization.
}
The objective of this paper is \PH{thus} to highlight effects of porosity variation in a topology optimization process in the field of fluid dynamics. 
To this end, we have developed in Section \ref{Governing equations} an original one-domain macroscopic model for the modelling of flow through variable porous media. Transition between fluid and porous domains is based on porosity $\varepsilon$ where we have the Navier-Stokes equation for $\varepsilon=1$ and a modified Darcy-Brinkman-Forchheimer equation for $\varepsilon \neq 1$. 
\newline
\PH{In Section \ref{adjoint method}, we begin by stating a topology optimization problem for porous media for a general cost function. Since solving numerically such problems with (e.g.) gradient-based algorithms needs the porosity to be smooth enough to be differentiated, we introduce a regularized version of our model from Section \ref{Governing equations}.
} 
We \PH{then validate both our model and our code} by reproducing numerical results from the literature, namely on the case of a porous plug problem where $\varepsilon$ varies from $1$ to $0.7$ and then goes back to $1$. 
\PH{We end this section by computing the continuous adjoint model used to get the gradient of a general cost functional.} 
\newline
Afterwards, in Section \ref{sec:results_num}, we perform topology optimization for two geometrical configurations \PH{that are classical in the topology optimization} literature aiming at minimizing the power dissipated by the fluid in the computational domain. This paper then ends with some conclusions and future works.

%


\section{Governing equations}
\label{Governing equations}

This section starts with the definition of the volume averaging method used to obtain a flow model suitable for a medium who presents 
\PH{spatially varying}
porosity. It consists of spatial averaging of phase behaviors over an elementary volume. The averaging volume, denoted by $V$ refers to a representative elementary volume (REV) which is occupied by a persistent solid phase $s$, with volume $V_s$ and void space occupied by fluid phase $f$ with volume $V_f$. \PH{The porosity $\varepsilon$ is then defined as: 
$$
\varepsilon=\frac{V_s}{V_s+V_f}=\frac{V_s}{V}.
$$
} 
The porosity is allowed to change between elementary volumes. In doing so, we are going to identify porosity coefficient inside gradient operators and thus gradient of porosity is taken into account in a modified Darcy-Brinkman-Forchheimer equation \PH{where we recover the Navier-Stokes equation for $\varepsilon=1$.} We end this section by estimating orders of magnitude of each term to obtain a reduced form of momentum conservation equation in the case of a convective flow. 

\subsection{Volume averaging}

\begin{figure}
    \centering
    \includegraphics[scale=0.35]{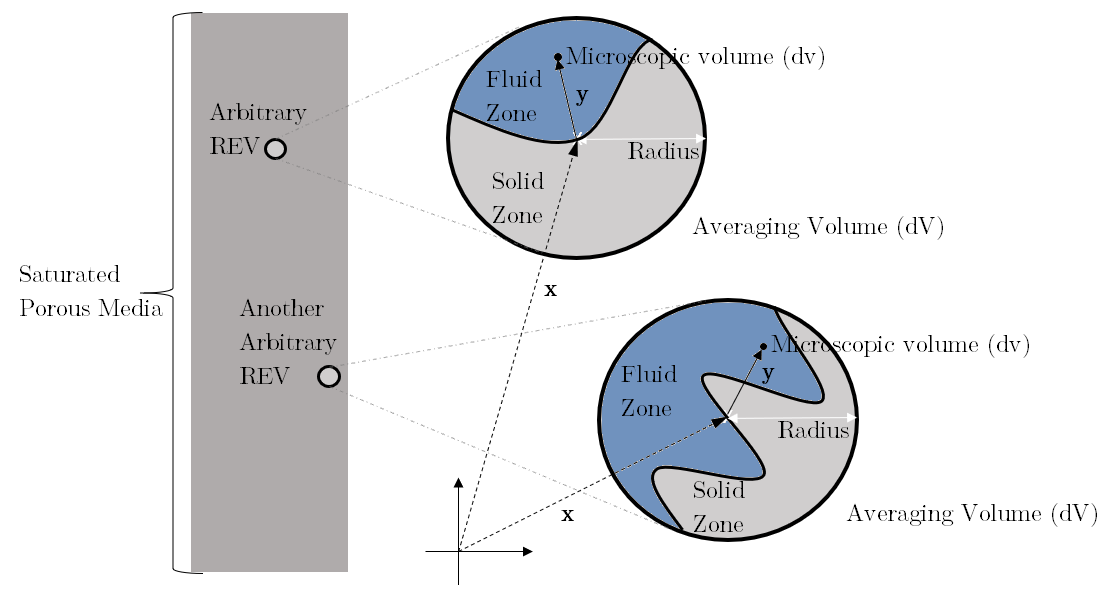}
    \caption{Representative elementary volumes}
    \label{rev}
\end{figure}

We use \cite[Appendix A]{BousquetMelou2002} to summarize the volume-averaging technique. A physical property $\psi$ is considered continuous in the phase fluid $f$ where $\psi=\psi_f$ and in the solid phase $s$ where $\psi=\psi_s$. The technique consists in averaging $\psi$ in a REV. 
Averages are calculated at the centroid of an averaging volume designated by the variable $\textbf{x}$ illustrated in Figure \ref{rev}. 
The following notations are adopted \cite[p. 12, Eq. (A8),(A9)]{BousquetMelou2002}: 
\begin{eqnarray*}
&& \langle \psi_f \rangle =\langle \psi_f \rangle\mid_{\textbf{x}}
=\frac{1}{V}\int_{V_s}\psi_s(\textbf{x}+\textbf{y}_s)\; dV_\textbf{y}
,
\\
&& \langle \psi_f \rangle^f 
=\langle \psi_f \rangle^f \mid_{\textbf{x}}
=\frac{1}{V_s}\int_{V_s}\psi_s(\textbf{x}+\textbf{y}_s)\; dV_\textbf{y},
\end{eqnarray*}
where $\langle \psi_f \rangle$ is superficial volume average and $\langle \psi_f \rangle^f$ is intrinsic phase average. \PH{We emphasize} that \PH{both these quantities actually depend on $\textbf{x}$} and are related by: 
 $$
 \langle \psi_f \rangle = \varepsilon \; \langle \psi_f \rangle^f.
 $$
\PH{In the sequel, we drop the braket and simply refer to $\psi^f$ to denote the intrinsic phase average of $\psi$.}


In this paper, problems are modeled under the assumptions of steady-state, Newtonian, incompressible and laminar flow in forced convection. We also assumed the fluid has constant density $\rho$ and constant viscosity $\mu$.

Following development in \cite[p. 71, Eq.(3.48)]{Diersch2014} \PH{and assuming that there is no phase-internal supply nor phase-change of mass}, the fluid mass conservation is: 
$$
\nabla \cdot (\varepsilon\;\rho\;\textbf{v}^f)=\PH{\rho}\nabla \cdot (\varepsilon\;\textbf{v}^f) =0,
$$ 
where $\textbf{v}^f$ is the particle velocity vector. Introducing the macroscopic velocity $\textbf{u}=\varepsilon\;\textbf{v}^f$, the fluid mass conservation reduces to:
$$\boxed{
\nabla \cdot \textbf{u} =0
}\ ,$$
\PH{which thus represents the incompressibility condition.}

For the fluid momentum conservation, we use the following relationships: 
\begin{itemize}
    \item \cite[p. 86, Eq. (3.114)]{Diersch2014}: 
$$\sigma^f=\mathrm{p}^f\delta + \tau^f,\ \varepsilon\;\rho\;\mathrm{f}^{f}_{\sigma} = \mathrm{p}^f\; \nabla \varepsilon + \mathrm{f}^{f}_{\tau},$$
    \item \cite[Eq. (3.142) p.92,]{Diersch2014}: 
    $$\tau^f = \dfrac{2}{3}\;\mu\;(\delta :\mathrm{\textbf{d}}^{\mathrm{f}})\delta-2\;\mu\;\mathrm{\textbf{d}}^{\mathrm{f}},$$ 
    \item \cite[Eq. (3.154) p.94]{Diersch2014}: 
    $$\mathrm{f}^{f}_{\tau} = - \varepsilon^2\mu \textbf{k}^{-1}\cdot \textbf{v}^{fs}-\varepsilon^3\rho\;\textbf{k}^{-1/2}\;\mathrm{c}_{\mathrm{F}}\;|\textbf{v}^{fs}|\cdot \textbf{v}^{fs},$$
\end{itemize}
where $\sigma^{f}$ is the stress tensor of the fluid, $\mathrm{p}^f$ is the thermodynamic pressure of the fluid, $\delta$ is the Kronecker symbol, $\tau^f$ is the deviatoric fluid stress tensor, $\mathrm{f}^{f}_{\sigma}$ is the interfacial drag term, $\mathrm{f}^{f}_{\tau}$ is the deviatoric fluid momentum exchange vector, $\mathrm{\textbf{d}}^{\mathrm{f}}$ is the rate of deformation tensor of the fluid phase with $\mathrm{\textbf{d}}^{\mathrm{f}}=\frac{1}{2}\left[\nabla \textbf{v}^f+(\nabla \textbf{v}^f)^\mathrm{T}\right]$ \PH{and $\textbf{k}$ is permeability tensor at full saturation}. 
\PH{We also recall that $\mathrm{c}_{\mathrm{F}}$ is the Forchheimer dimensionless form-drag constant and note that it is often approximated by $0.55$. In addition,$ \textbf{v}^{fs}$ is the relative velocity defined as: 
$$
\textbf{v}^{fs}=\textbf{v}^f-\textbf{v}^s,
$$
and since the velocity inside the solid is $\textbf{v}^s=\textbf{0}$, we get 
$\textbf{v}^{fs}=\textbf{v}^f$.
}
Now assuming that:
\begin{itemize}
    \item external supply of momentum $\mathrm{g}^f=0$,
    \item dynamic viscosity $\mu$ is constant, 
    \item $\textbf{v}^f$ is solenoidal and $\nabla \cdot \textbf{v}^f = \delta : \mathrm{\textbf{d}}^{\mathrm{f}}=0$,
\end{itemize}
the momentum conservation equation \PH{(see \cite[p. 73, Eq. (3.56)]{Diersch2014}) becomes}: 
\begin{equation}
\begin{array}{ll}
\label{eq3}
\nabla \cdot (\varepsilon\;\rho\;\textbf{v}^{f}\;\textbf{v}^{f}) + \nabla (\varepsilon\;\mathrm{p}^f) - 2\;\mu\; \nabla \cdot (\varepsilon\;\mathrm{\textbf{d}}^{\mathrm{f}})\\[5mm]=\mathrm{p}^f\; \nabla \varepsilon - \varepsilon^2\mu \textbf{k}^{-1}\cdot \textbf{v}^{f}-\varepsilon^3\rho\;\textbf{k}^{-1/2}\;\mathrm{c}_{\mathrm{F}}\;|\textbf{v}^{f}|\cdot \textbf{v}^{f}.
\end{array}
\end{equation}
Introducing the macroscopic velocity $\textbf{u}=\varepsilon\;\textbf{v}^f$ and using $\nabla (\varepsilon\;\mathrm{p}^f)=\varepsilon \nabla \mathrm{p}^f + \mathrm{p}^f \nabla \varepsilon$, \PH{Eq. (\ref{eq3}) can be recast as follow}:
\begin{equation}
\label{eq:Momentum_almost_done}
\begin{array}{ll}
\nabla \cdot (\frac{\rho}{\varepsilon}\;\textbf{u}\;\textbf{u}) + \varepsilon \nabla \mathrm{p}^f - 2\;\mu\; \nabla \cdot (\varepsilon\;\mathrm{\textbf{d}}^{\mathrm{f}})\\[5mm]=- \varepsilon \;\mu \textbf{k}^{-1}\cdot \textbf{u}-\varepsilon\;\rho\;\textbf{k}^{-1/2}\;\mathrm{c}_{\mathrm{F}}\;|\textbf{u}|\cdot \textbf{u}.
\end{array}
\end{equation}
\PH{We now wish to expand} the term with the rate of 
deformation tensor \PH{from Eq. (\ref{eq:Momentum_almost_done})}. This gives:
\begin{equation}
\label{eq:Deformation_tensor_1}
\begin{array}{lll}
- 2\;\mu\; \nabla \cdot (\varepsilon\;\mathrm{\textbf{d}}^{\mathrm{f}}) \\[5mm]= - \varepsilon\;\mu\;\Delta \textbf{v}^{\mathrm{f}}-\mu\;(\nabla \textbf{v}^{\mathrm{f}}+\left[\nabla \textbf{v}^{\mathrm{f}}\right]^{\mathrm{T}}) \cdot \nabla \varepsilon \\[5mm]=- \varepsilon\;\mu\;\Delta (\dfrac{\textbf{u}}{\varepsilon}) - \mu\;\nabla (\dfrac{\textbf{u}}{\varepsilon})\cdot \nabla \varepsilon - \mu\;\left[\nabla (\dfrac{\textbf{u}}{\varepsilon})\right]^{\mathrm{T}}\cdot \nabla \varepsilon,
\end{array}
\end{equation}
\PH{where we used}:
\begin{equation}
\label{eq:Deformation_tensor_2}
\begin{array}{ll}
\Delta (\dfrac{\textbf{u}}{\varepsilon}) = \dfrac{1}{\varepsilon} \Delta \textbf{u} + \textbf{u} \Delta (\dfrac{1}{\varepsilon}) + 2 \nabla \textbf{u} \cdot \nabla (\dfrac{1}{\varepsilon}) \\[5mm]= \dfrac{1}{\varepsilon} \Delta \textbf{u} - \dfrac{\textbf{u}}{\varepsilon^{2}} \Delta \varepsilon + \dfrac{2\textbf{u}}{\varepsilon^{3}}|\nabla \varepsilon|^{2} - \dfrac{2}{\varepsilon^{2}}\nabla \textbf{u} \cdot \nabla \varepsilon,
\end{array}
\end{equation}
\PH{to simplify it further}. For the convective term \PH{from Eq. (\ref{eq:Momentum_almost_done})}, we have:
\begin{equation}
\label{eq:convective_term}
\rho\;\nabla \cdot (\dfrac{1}{\varepsilon}\;\textbf{u}\;\textbf{u}) = \dfrac{\rho}{\varepsilon}\;\nabla \cdot (\textbf{u}\;\textbf{u}) - \dfrac{\rho}{\varepsilon^{2}}\;(\textbf{u} \; \textbf{u})\cdot \nabla \varepsilon.
\end{equation}
\PH{Now gathering (\ref{eq:Deformation_tensor_1},\ref{eq:Deformation_tensor_2}) and (\ref{eq:convective_term}),
Eq. (\ref{eq:Momentum_almost_done}) becomes:
} 
$$
\begin{array}{lll}
\dfrac{\rho}{\varepsilon}\;\nabla \cdot (\textbf{u}\;\textbf{u}) - \dfrac{\rho}{\varepsilon^{2}}\;(\textbf{u} \; \textbf{u})\cdot \nabla \varepsilon + \varepsilon \nabla \mathrm{p}^f \\[5mm]- \varepsilon\;\mu\;\left(\dfrac{1}{\varepsilon} \Delta \textbf{u} - \dfrac{\textbf{u}}{\varepsilon^{2}} \Delta \varepsilon + \dfrac{2\textbf{u}}{\varepsilon^{3}}|\nabla \varepsilon|^{2} - \dfrac{2}{\varepsilon^{2}}\nabla \textbf{u} \cdot \nabla \varepsilon \right)
\\[5mm]- \mu\;\nabla (\dfrac{\textbf{u}}{\varepsilon})\cdot \nabla \varepsilon - \mu\;\left[\nabla (\dfrac{\textbf{u}}{\varepsilon})\right]^{\mathrm{T}}\cdot \nabla \varepsilon\\[5mm]=- \varepsilon \;\mu \textbf{k}^{-1}\cdot \textbf{u}-\varepsilon\;\rho\;\textbf{k}^{-1/2}\;\mathrm{c}_{\mathrm{F}}\;|\textbf{u}|\cdot \textbf{u},
\end{array}
$$
\PH{from which} the momentum conservation equation can be finally written as:
\begin{equation}
\label{eq:Final_momentum}
\begin{array}{lll}
\dfrac{\rho}{\varepsilon}\;\nabla \cdot (\textbf{u}\;\textbf{u}) + \varepsilon \nabla \mathrm{p}^f - \mu \; \Delta \textbf{u} \\[5mm]
- \left( \dfrac{ \rho}{\varepsilon^{2}}\;(\textbf{u} \; \textbf{u}) - \dfrac{2\;\mu}{\varepsilon}\nabla \textbf{u}+ \mu\;\nabla (\dfrac{\textbf{u}}{\varepsilon}) + \mu\;\left[\nabla (\dfrac{\textbf{u}}{\varepsilon})\right]^{\mathrm{T}} \right) \cdot \nabla \varepsilon \\[5mm]+ \dfrac{\mu \textbf{u}}{\varepsilon} \Delta \varepsilon - \dfrac{2\mu \textbf{u}}{\varepsilon^{2}}|\nabla \varepsilon|^{2} 
+\varepsilon\; \mu \; \mathrm{\textbf{k}}^{-1}\cdot \textbf{u} + \varepsilon\; \rho \; \mathrm{\textbf{k}}^{-1/2} \; \mathrm{c}_{\mathrm{F}}\;\textbf{u}\cdot |\textbf{u}|\\[5mm]=\textbf{0},
\end{array}
\end{equation}
\PH{where the contribution of the variation/gradient of porosity clearly appears.}

\PH{To get Eq. (\ref{eq:Final_momentum})} 
in dimensionless form, we use the following transformations: 
$$
\textbf{u}^*=\dfrac{\textbf{u}}{U},\ (\textbf{x}^*,\textbf{y}^*,\textbf{z}^*)=
\dfrac{(\textbf{x},\textbf{y},\textbf{z})}{L},\  \left(\mathrm{p}^f\right)^*=\dfrac{\mathrm{p}^f}{\rho\;U^2}.
$$
The differential operators in the dimensionless coordinates system are related 
to the original ones through: 
$$
\nabla^*(\bullet)=L\;\nabla(\bullet),\ \Delta^*(\bullet)=L^2\;\Delta(\bullet),
$$ 
and the dimensionless form of (\ref{eq:Final_momentum}) then reads:
\begin{equation}
\label{dimensionless}
\begin{array}{lll}
\nabla^* \cdot (\dfrac{1}{\varepsilon}\;\textbf{u}^*\;\textbf{u}^*) + \varepsilon \nabla^* (\mathrm{p}^f)^* - \nabla^* \cdot(\dfrac{1}{\mathrm{Re}}\;\nabla^* \textbf{u}^*) \\[5mm]

+\left( \dfrac{2}{\varepsilon\;\mathrm{Re}}\nabla^* \textbf{u}^* - \dfrac{1}{\mathrm{Re}}\;\nabla^* (\dfrac{\textbf{u}^*}{\varepsilon}) - \dfrac{1}{\mathrm{Re}}\;\left[\nabla^* (\dfrac{\textbf{u}^*}{\varepsilon})\right]^{\mathrm{T}}\right) \cdot \nabla^* \varepsilon \\[5mm]+ \dfrac{1}{\varepsilon\;\mathrm{Re}}\;\textbf{u}^*\; \Delta^* \varepsilon - \dfrac{2}{\varepsilon^{2}\;\mathrm{Re}}\;\textbf{u}^*\;|\nabla^* \varepsilon|^{2} \\[5mm]

+ \dfrac{\varepsilon}{\mathrm{Re}\;\mathrm{Da}\;\dfrac{\mathrm{\textbf{k}}(\varepsilon)}{\mathrm{\textbf{k}}_{0}}}\cdot \textbf{u}^* + \dfrac{\varepsilon\;\mathrm{c}_{\mathrm{F}}}{\left(\mathrm{Da}\;\dfrac{\mathrm{\textbf{k}}(\varepsilon)}{\mathrm{\textbf{k}}_{0}}\right)^{1/2}} \;\textbf{u}^*\cdot |\textbf{u}^*|=\textbf{0},
\end{array}
\end{equation}
where we also introduced the Reynolds and Darcy numbers, respectively defined as: 
$$
\mathrm{Re}=\dfrac{V \rho L}{\mu},\ \mathrm{Da}=\dfrac{\mathrm{\textbf{k}}_{0}}{L^{2}},$$ 
with $V$ a characteristic velocity and $L$ a characteristic length. 
We finally define the remaining physical constants from (\ref{dimensionless}).
The permability is defined using the Carman-Kozeny relationship \cite{Amiri1994,Kruczek2014} and reads: 
$$
\mathrm{\textbf{k}}=\mathrm{\textbf{k}}(\varepsilon)=\dfrac{\varepsilon^{3}\;\mathrm{d}^{2}_{\mathrm{p}}}{K(1-\varepsilon)^{2}},
$$
with $K$ an empirical constant and $\mathrm{d}_{\mathrm{p}}$ the mean particle diameter. The Darcy and Forchheimer terms are rewritten by introducing a coefficient: 
$$
\beta = \dfrac{K\;L^{2}}{\mathrm{d}^{2}_{\mathrm{p}}},
$$
calculated with the characteristics of the porous media. 
In doing so and having in mind that all quantities are without dimension\PH{, we can drop the $*$ in (\ref{dimensionless}) for clarify equations and }
 the final momentum conservation equation can be written:
\begin{equation}
\label{eqbase}
\boxed{
\begin{array}{lll}
\nabla \cdot (\dfrac{1}{\varepsilon^2}\;\textbf{u}\;\textbf{u}) + \nabla \mathrm{p}^f - \nabla \cdot(\dfrac{1}{\varepsilon\;\mathrm{Re}}\;\nabla \textbf{u}) +\dfrac{1}{\varepsilon^3}\;\textbf{u}\;\textbf{u} \cdot \nabla \varepsilon
\\[5mm]
+\left(\dfrac{1}{\varepsilon^2\;\mathrm{Re}}\nabla \textbf{u} - \dfrac{1}{\varepsilon\;\mathrm{Re}}\;\nabla (\dfrac{\textbf{u}}{\varepsilon}) - \dfrac{1}{\varepsilon\;\mathrm{Re}}\;\left[\nabla (\dfrac{\textbf{u}}{\varepsilon})\right]^{\mathrm{T}}\right) \cdot \nabla \varepsilon 
\\[5mm]
+ \dfrac{1}{\varepsilon^2\;\mathrm{Re}}\;\textbf{u}\; \Delta \varepsilon - \dfrac{2}{\varepsilon^{3}\;\mathrm{Re}}\;\textbf{u}\;|\nabla \varepsilon|^{2} + \beta \;\dfrac{(1-\varepsilon)^{2}}{\varepsilon^{3}\;\mathrm{Re}}\;\textbf{u} 
\\[5mm]
+ \beta^{1/2} \; \dfrac{\mathrm{c}_{\mathrm{F}}\;(1-\varepsilon)}{\varepsilon^{3/2}} \;\textbf{u}\cdot |\textbf{u}|=\textbf{0}.
\end{array}}
\end{equation} 
We end with Eq. (\ref{eqbase}) which is an original macroscopic model of momentum conservation in a fluid through porous matrix. We also identify Brinkman-like terms as pointed by \cite{Goyeau1997}. Although terms of gradient of porosity that follow on taking account on porosity variation can be found in literature, our approach stands out. We proceed from a microscopic balance equation for a physical property $\psi$ in following development in \cite{Diersch2014} and introduce relationships and assumptions that permit closure of momentum equation at the outset. In addition, Darcy and Forchheimer terms rewritten in introducing Carman-Kozeny relationship and $\beta$ coefficient let us study explicitely effect of particle diameter.

\subsection{Model reduction}
\label{model reduction}
Our model presented in Eq. (\ref{eqbase}) contains many terms that involve Reynolds number. We now estimate each terms in order to obtain a reduced model for a convective flow. To this end, we use the reference \cite{Goyeau1997} which established that average properties of evolving heterogeneities in porous matrix are not only point dependent but also depend on the size of the averaging volume. According to this statement, the previous authors proposed the following length scale constraint:  
$$
l_\beta \leqslant r_0 \sim L_\varepsilon, L_\textbf{u},
$$
where $l_\beta$ is a pore length scale, $r_0$ is the radius of averaging volume, $L_\varepsilon$ and $L_\textbf{u}$ are respectively characteristic length for porosity and averaged velocity. We note by $\bigtriangleup$ the variation of a quantity and assume that: 
$$
\bigtriangleup \varepsilon = \mathcal{O} \left(1\right),\ 
\textbf{u} = \mathcal{O} \left(\mathrm{U}\right).
$$ 
\PH{Note that this ensures}:
$$
\bigtriangleup \textbf{u} = \mathcal{O} \left(\mathrm{U}\right).
$$
We see in \cite{Whitaker1999} that: 
$$\nabla \phi = \mathcal{O} \left(\bigtriangleup \phi / L_\phi\right),
$$ 
where $\phi$ is quantity of interest and $L_\phi$ the associated characteristic length.
\PH{We emphasize that all quantities, namely $U,L_\varepsilon, L_\textbf{u}$, are all dimensionless.}
We have: 
$$
\nabla\left(\dfrac{\textbf{u}}{\varepsilon}\right)=\dfrac{\nabla \textbf{u}}{\varepsilon} + \textbf{u}\;\nabla \left(\dfrac{1}{\varepsilon}\right)=\dfrac{\nabla \textbf{u}}{\varepsilon}- \dfrac{\textbf{u}\;\nabla \varepsilon}{\varepsilon^2}= \mathcal{O} \left(\dfrac{U}{\varepsilon\;L_\textbf{u}}-\dfrac{U}{\varepsilon^2\;L_\varepsilon}\right).
$$
Considering then that: 
$$
L_\textbf{u} \sim L_\varepsilon = L_0,
$$ 
we have: $$\nabla\left(\dfrac{\textbf{u}}{\varepsilon}\right)=\mathcal{O} \left( \dfrac{U}{\varepsilon^2\;L_0}\right).
$$ 
\PH{Similar computations show that} the order of magnitude of the convective term is:
$$
\begin{array}{lll}
\nabla \cdot \left(\dfrac{1}{\varepsilon^2}\;\textbf{u}\;\textbf{u}\right)= \dfrac{1}{\varepsilon^2}\;\nabla \cdot (\textbf{u}\;\textbf{u}) + \textbf{u}\;\textbf{u} \cdot \nabla (\dfrac{1}{\varepsilon^2})\\[5mm]= \dfrac{1}{\varepsilon^2}\;\nabla \cdot (\textbf{u}\;\textbf{u}) - \dfrac{2}{\varepsilon^3} \textbf{u}\;\textbf{u} \cdot \nabla \varepsilon = \mathcal{O} \left(\dfrac{U^2}{\varepsilon^3\;L_0}\right),
\end{array}
$$
and that the following terms are of the same order:
$$
\begin{array}{lll}
\dfrac{1}{\varepsilon^3}\;\textbf{u}\;\textbf{u} \cdot \nabla \varepsilon = \mathcal{O} \left(\dfrac{U^2}{\varepsilon^3\;L_0}\right), 
\end{array}
$$


for the terms involving $\beta$, one has: 
$$
\begin{array}{rll}
\beta^{1/2} \; \dfrac{\mathrm{c}_{\mathrm{F}}\;(1-\varepsilon)}{\varepsilon^{3/2}} \;\textbf{u}\cdot |\textbf{u}| &=& \mathcal{O}\left(\sqrt{\beta}\dfrac{\mathrm{c}_{\mathrm{F}}\;(1-\varepsilon)}{\varepsilon^{3/2}} \;U^2 \right) \\
&=&\mathcal{O}\left(\sqrt{\beta}L_0\left(\dfrac{U^2}{\varepsilon^3\;L_0}\right)\right),
\\
\beta \;\dfrac{(1-\varepsilon)^{2}}{\varepsilon^{3}\;\mathrm{Re}}\;\textbf{u} 
&=& 
\mathcal{O}\left( \beta \;\dfrac{(1-\varepsilon)^{2}}{\varepsilon^{3}\;\mathrm{Re}}\;U\right) \\
&=&\mathcal{O}\left( \dfrac{\beta\; L_0}{U\; \mathrm{Re}}\left(\dfrac{U^2}{\varepsilon^{3} L_0}\right)\right).
\end{array}
$$
Assuming then $\beta,\ U,\ L_0,\ \mathrm{Re}$ satisfy for instance:
\begin{equation}\label{eq:hyp_beta}
\sqrt{\beta}L_0\geq 1,\ U\geq 1,\ \dfrac{\beta\;L_0}{\mathrm{Re}}\geq 1,
\end{equation}
we obtain that the Forchheimer and Darcy terms are of the same magnitude as the convective term. Note that assumptions (\ref{eq:hyp_beta}) only ensure that these parameters are not too small.

\PH{Conducting a similar analysis, we can estimate the terms below}: 
$$
\begin{array}{lll}
 \nabla \cdot(\dfrac{1}{\varepsilon\;\mathrm{Re}}\;\nabla \textbf{u})=\dfrac{1}{\varepsilon^2\;\mathrm{Re}}\left(\varepsilon\Delta \textbf{u} - \nabla \textbf{u}\;\nabla \varepsilon\right)=\mathcal{O} \left(\dfrac{1}{\varepsilon^2\;\mathrm{Re}} \dfrac{U}{L_0^2}\right), 
\end{array}
$$

$$
\begin{array}{lll}
\displaystyle \dfrac{1}{\varepsilon^2\;\mathrm{Re}}\nabla \textbf{u} \cdot \nabla \varepsilon= \mathcal{O} \left(\frac{1}{\varepsilon^2\;\mathrm{Re}}\frac{\mathrm{U}}{L_0^2}\right), 
\end{array}
$$

$$
\begin{array}{lll}
\dfrac{1}{\varepsilon\;\mathrm{Re}}\;\nabla (\dfrac{\textbf{u}}{\varepsilon}) \cdot \nabla \varepsilon
\displaystyle=\mathcal{O}\left(\frac{1}{\varepsilon^3\;\mathrm{Re}}\frac{\mathrm{U}}{L_0^2}\right), 
\end{array}
$$

$$
\begin{array}{lll}
 \displaystyle\dfrac{1}{\varepsilon^2\;\mathrm{Re}}\;\textbf{u}\; \Delta \varepsilon=\mathcal{O}\left(\frac{1}{\varepsilon^2\;\mathrm{Re}}\frac{\mathrm{U}}{L_0^2}\right), 
 \end{array}
$$

$$
\begin{array}{lll}
\displaystyle\dfrac{2}{\varepsilon^{3}\;\mathrm{Re}}\;\textbf{u}\;|\nabla \varepsilon|^{2}=\mathcal{O}\left(\frac{2}{\varepsilon^{3}\;\mathrm{Re}}\frac{\mathrm{U}}{L_0^2}\right). 
 \end{array}
$$
\PH{We emphasize that all the previous terms are almost of the same magnitude and we now show that they are actually negligeable. First, note that since $\varepsilon\leq 1$, we 
have: 
$$
\dfrac{1}{\varepsilon^2\;\mathrm{Re}} \dfrac{U}{L_0^2}\leq \dfrac{1}{\varepsilon^3\;\mathrm{Re}} \dfrac{U}{L_0^2},
$$
if the Renoylds number and $L_0$ are such that:
$$
L_0\mathrm{Re}\gg 1,
$$
one can infer: 
$$
\dfrac{1}{\varepsilon^3\;\mathrm{Re}} \dfrac{U}{L_0^2}\ll \dfrac{U}{\varepsilon^3\;L_0}\leq 
\dfrac{U^2}{\varepsilon^3\;L_0},
$$
and the terms below:
\begin{eqnarray*}
&&\nabla \cdot(\dfrac{1}{\varepsilon\;\mathrm{Re}}\;\nabla \textbf{u})\textbf{;}
\
\dfrac{1}{\varepsilon^2\;\mathrm{Re}}\nabla \textbf{u} \cdot \nabla \varepsilon\textbf{;}
\\
&& \dfrac{1}{\varepsilon\;\mathrm{Re}}\;\nabla (\dfrac{\textbf{u}}{\varepsilon}) \cdot \nabla \varepsilon\textbf{;}
\, \dfrac{1}{\varepsilon^2\;\mathrm{Re}}\;\textbf{u}\; \Delta \varepsilon \textbf{;} \, \dfrac{2}{\varepsilon^{3}\;\mathrm{Re}}\;\textbf{u}\;|\nabla \varepsilon|^{2},
\end{eqnarray*}
can thus be disregarded from Eq. (\ref{eqbase}).
}

\PH{To summarize the analysis conducted in this section so far, if we assume that}:
\begin{eqnarray*}
&& L_\textbf{u} \sim L_\varepsilon = L_0,\;
U\geq 1,\;
\sqrt{\beta}L_0\geq 1,\\
&& \dfrac{\beta\;L_0}{\mathrm{Re}}\geq 1 \text{ and }
L_0\mathrm{Re}\gg 1,
\end{eqnarray*}
we obtain a reduced form of Eq. (\ref{eqbase}) in maintaining the Brinkman term:
\begin{equation}
\label{modelsimplified}
\boxed{
\begin{array}{lll}
\nabla \cdot (\dfrac{1}{\varepsilon^2}\;\textbf{u}\;\textbf{u}) +\dfrac{1}{\varepsilon^3}\;\textbf{u}\;\textbf{u} \cdot \nabla \varepsilon + \nabla \mathrm{p}^f - \nabla \cdot(\dfrac{1}{\varepsilon\;\mathrm{Re}}\;\nabla \textbf{u}) \\[5mm] +\beta \;\dfrac{(1-\varepsilon)^{2}}{\varepsilon^{3}\;\mathrm{Re}}\;\textbf{u}+ \beta^{1/2} \; \dfrac{\mathrm{c}_{\mathrm{F}}\;(1-\varepsilon)}{\varepsilon^{3/2}} \;\textbf{u}\cdot |\textbf{u}|=\textbf{0}.
\end{array}}
\end{equation}
Eq. (\ref{modelsimplified}) is interesting because porosity coefficients are effectively inside gradient operators for the convective term and the Brinkman term when porosity varies. In addition to this, there is a source term $(\dfrac{1}{\varepsilon^3}\;\textbf{u}\;\textbf{u} \cdot \nabla \varepsilon)$ which is new and who appears when porosity variation \PH{have been taken into account.} That make Eq. (\ref{modelsimplified}) different from the usual Darcy-Brinkman-Forchheimer equation and suitable for a variable porosity medium in the case of convective flow.

\section{Topology optimization problem}
\label{adjoint method}
\PH{In the previous section, we derived a mathematical model for steady state incompressible flow in variable porous media.}
We now wish to perform topology optimization \PH{using Eq. (\ref{modelsimplified}) as constraints. 
We are then going to consider the porosity $\varepsilon$ as a function of a design variable $\alpha$. The fluid zones can then be defined as: 
$$
\Omega_f:=\left\{x\in\Omega\ |\ \varepsilon(\alpha(x))=1\right\},
$$
while the porous zones with a given constant porosity $\varepsilon_0$ 
are:
$$
\Omega_p:=\left\{x\in\Omega\ |\ \varepsilon(\alpha(x))=\varepsilon_0 \right\}.
$$
We emphasize that such function taking only two discrete values leads to optimization problems that are difficult or even impossible to solve 
(see e.g. \cite{evgrafov2005limits}).} 
As it is usually done in topology optimization, we thus introduce a regularisation function used to interpolate porosity by means of the design variable $\alpha$. Such regularization also allows to use gradient-based algorithms to numerically solve the topology optimization problem.
\PH{We adopt the methodology from} \cite{Ramalingom2018}
and then use:
\begin{equation}
\label{sigmoid2}
\varepsilon(\alpha)=(\varepsilon_{0}-\varepsilon_{f})h_\tau(\alpha)+\varepsilon_{f},
\end{equation}
where: 
$$
h_\tau(\alpha)=\left(\dfrac{1}{1+\exp{(-\tau(\alpha-\alpha_{0}))}}-\dfrac{1}{1+\exp{(\tau\;\alpha_{0})}}\right),
$$
is a smooth regularization of a step function which, as $\tau\to+\infty$, goes to $0$ for $\alpha<\alpha_0$ and to $1$ for $\alpha>\alpha_0$. 
In addition, one have $\alpha \in \left[0,\; \alpha_{max} \right]$ \PH{for a given} $\alpha_{max}$ and $\varepsilon_f=1$. \PH{We note that the fluid zones $\Omega_f$ are obtained, as $\tau\to+\infty$, when $\alpha\leq \alpha_0$ and the porous zones $\Omega_p$ when $\alpha>\alpha_0$.}
\PH{It is worth noting that} regularising the porosity $\varepsilon$ allows to correctly define it's derivative. \PH{The latter is of great help} since gradient of porosity are taking account in our flow model (\ref{modelsimplified}).

The topology optimization problem considered in this paper is finally written in the following general form:
\begin{equation}
\label{eq:TO_problem}
\begin{array}{lll}
\text{Minimize}& \mathrm{J}\text{ (}\textbf{u}\text{, } \mathrm{p}\text{, } \alpha\text{)}\\
\text{Subject to} & \text{Equations}\ (\ref{modelsimplified})\ \text{for} \text{ (}\textbf{u}\text{, } \mathrm{p}\text{, } \alpha\text{)}\\
& \text{Boundary conditions on }\Gamma.
\end{array}
\end{equation}

Assuming that $\partial\Omega=\Gamma=\Gamma_{\mathrm{in}}\cup\Gamma_{\mathrm{w}}\cup\Gamma_{\mathrm{out}}$ is decomposed respectively with an inlet, walls and an outlet, we use the next boundary conditions for the direct problem:
\begin{eqnarray}
\nonumber && \mathrm{on\ }\Gamma_{\mathrm{in}}:\ \textbf{u}=\textbf{u}_{\mathrm{in}}=1,\ \partial_\textbf{n}\mathrm{p}=0,  \\
&& \mathrm{on\ }\Gamma_{\mathrm{w}}:\ \textbf{u}=0,\  \partial_\textbf{n}\mathrm{p}=0, \\
\nonumber && \mathrm{on\ }\Gamma_{\mathrm{out}}:\ \partial_\textbf{n}\textbf{u}=0,\ \mathrm{p}=0.
\end{eqnarray}



In the sequel of this section, we first reproduce  a numerical case study of \cite{DeGroot2011} and \cite{Betchen2006}. These results are going to serve as validation of our code as well as the regularization of porosity introduced in our state problem (\ref{modelsimplified}). 
Next, we compute the adjoint model associated to our optimizarion problem (\ref{eq:TO_problem}) since the latter is going to be needed to compute the gradient of the cost function. 

\subsection{Justification of primal equations and code validation}
\label{porous plug flow}

We justify our flow models Eq. (\ref{eqbase}) and Eq. (\ref{modelsimplified}) in a two-dimensional rectangular geometry of length $L$ containing a porous insert in the interval $[\textbf{x}_1,\;\textbf{x}_2]$ while fluid zones are located in the inverval $[0,\;\textbf{x}_1]\cup[\textbf{x}_2,\;L]$. We chose to interpolate the porosity $\varepsilon$ with the help of the following regularisation function:
$$
\varepsilon(\textbf{x}) = S_1(\textbf{x}) \times S_2(\textbf{x}),
$$
with:
\begin{eqnarray}
\label{sigmoid1}
\nonumber\displaystyle    S_1(\textbf{x})&=& \varepsilon_{f}  \\
\nonumber & + & (\varepsilon_{0}-\varepsilon_{f})\left(\frac{\mathrm{tanh}(\frac{\tau(\textbf{x}-\textbf{x}_1)}{2})-\mathrm{tanh}(-\frac{\tau\;\textbf{x}_1}{2})}{2}\right),\\
 \displaystyle   S_2(\textbf{x})&=&\varepsilon_{f} \\
  \nonumber& + & (\frac{1}{\varepsilon_{0}}-\varepsilon_{f})\left(\frac{\mathrm{tanh}(\frac{\tau(\textbf{x}-\textbf{x}_2)}{2})-\mathrm{tanh}(-\frac{\tau\;\textbf{x}_2}{2})}{2}\right),
\end{eqnarray}
where $\varepsilon_{f}=1$ the fluid porosity, $\varepsilon_{0}$ the porous material porosity and $\tau$ a parameter who influences stiffness of the transition between fluid and porous zones. In this case, porosity interpolation \PH{hence variations of this quantity stem from a} physical basis. Indeed, porosity near interface region differs from that in the porous core, thus the porosity undergoes a spatial variation \PH{as illustrated in Figure 5 and 6 from} \cite{Khalili2014}. Our interpolation \PH{then} follows the same idea in considering a transition layer thickness by the parameter $\tau$. 

We reproduce a porous plug problem studied by \cite{DeGroot2011} and \cite{Betchen2006}. Parameters are: 
$$
\varepsilon=0.7,\ \mathrm{Re}_H=1,000,\ \mathrm{Da}=10^{-2},10^{-3},10^{-4},10^{-5}.
$$
It is a two dimensional \PH{rectangular} geometry \PH{with dimensions} $L \times H$. $L$ is composed of fluid parts $\Delta x_1$ and $\Delta x_3$ and porous part on $\Delta x_2$. We used a grid with grading in the flow direction and applied a cosinus function in the $H$ direction. To reproduce fully developed velocity boundary conditions at the inlet and at the outlet, we choosed $\Delta x_1=\Delta x_3=200H$. Pressure boundary condition is zero at the outlet and was extrapolated to all other boundaries. We applied a no slip condition at the walls. The porous insert has a length of $\Delta x_2=5H$. Numerical schemes used were Gauss cubic for gradient operators, Gauss upwind for convective term, Gauss linear for other divergen\PH{ce-like} terms and limited Gauss linear corrected $0.5$ for Laplacian terms. According to \cite[p. 4, Eq. (14)]{Kozelkov2018}, velocity and pressure on a face $f$ are calculated with opposite interpolation factors, we have: $\textbf{u}_f=(1-\lambda_f)\textbf{u}_p+\lambda_f\textbf{u}_N$ and $\mathrm{p}_f=\lambda_f \mathrm{p}_p+(1-\lambda_f)\mathrm{p}_n$ where $\lambda_f$ is the interpolation factor and $f$ is a face separating cells $P$ and $N$.

\begin{figure*}
    \centering
    \includegraphics[scale=0.4]{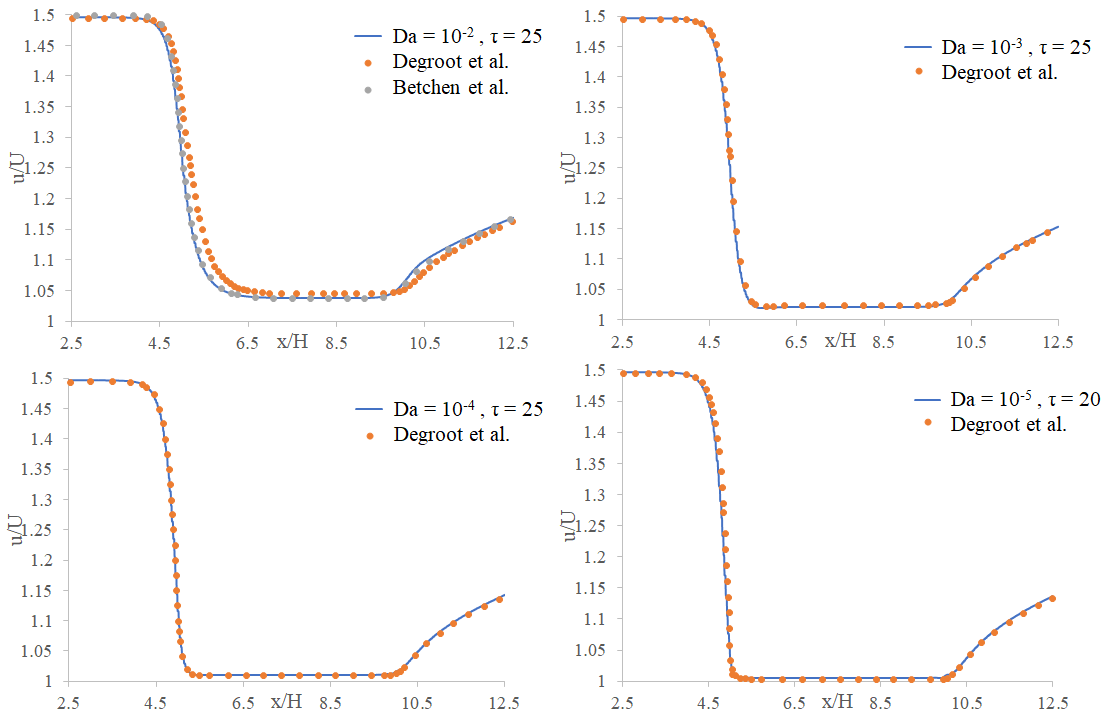}
    \caption{Results obtained in comparison with \cite{Betchen2006} and \cite{DeGroot2011}. Porous insert in the interval $[5,\; 10]$. $\mathrm{Da}=(10^{-2},\;10^{-3},\;10^{-4},\;10^{-5}$) and $\mathrm{Re}_H=1,000$.}
    \label{comparaison_degroot_betchen}
\end{figure*}

Results obtained are compared in terms of adimensional velocity magnitudes along the line $\mathrm{y}$/$\mathrm{H}$=0.5 on the interval $\mathrm{x}$/$\mathrm{H} \in [2.5,\;12.5]$ with results of \cite{DeGroot2011} and \cite{Betchen2006}. They are presented in Figure \ref{comparaison_degroot_betchen} for different Darcy numbers ($10^{-2}$, $10^{-3}$, $10^{-4}$ and $10^{-5}$).

\begin{figure}
    \centering
    \includegraphics[scale=0.5]{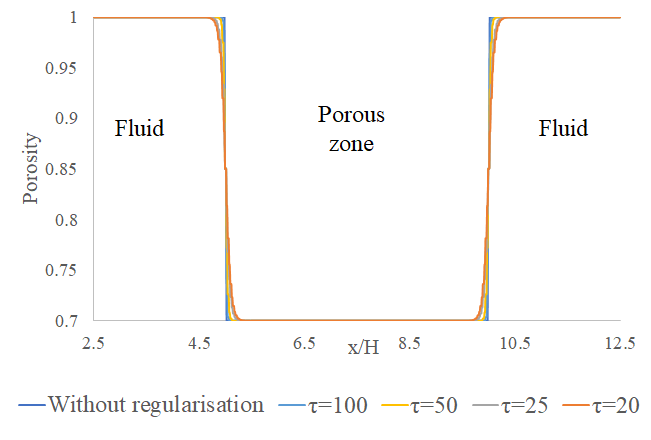}
    \caption{Influence of $\tau$ parameter in the transition between fluid and porous domains}
    \label{porosity_regularisation}
\end{figure}

Although results are presented for the complete model Eq. (\ref{eqbase}), it is interesting to note that the reduced model Eq. (\ref{modelsimplified}) allows getting exactly same outputs. We observe a very good agreement between our simulations results and results obtained in literature. For $\mathrm{Da}=10^{-2}$, velocity profile corresponds exactly to that obtained in Betchen et al. \cite{Betchen2006} and for $\mathrm{Da}=10^{-3}$, $10^{-4}$ and $10^{-5}$ to that obtained in Degroot et al. \cite{DeGroot2011}. There is a slight discrepancy for $\mathrm{Da}=10^{-2}$ due to different treatment of pressure at interfaces between fluid and porous domains \cite{DeGroot2011}. Either way, we managed to reproduce flow results in the transition between pure fluid and porous regions without the need of particular interfaces pressure conditions as done in \cite{DeGroot2011} and \cite{Betchen2006}. To \PH{do so,} we have chosen a continuous transition for porosity variation with the help of a sigmoid function (see Eq. (\ref{sigmoid1})). In the expression of this function, there is a particular $\tau$ number which made it possible to obtain consistent results. This $\tau$ value was found gradually and corresponds to the moment where there are no numerical oscillations in the velocity profile. This $\tau$ parameter influences stiffness of the \PH{fluid/porous transition zones} and simulations without regularisation (not presented here) resulted with non-physical oscillations that became more pronounced as Darcy number decreases. Non-physical oscillations comes from numerical procedure as pointed by \cite{DeGroot2011}. The effect of the regularisation function is illustrated in Figure \ref{porosity_regularisation} for different values of $\tau$. High value of $\tau$ results in a sharp transition and a low value creates a smooth transition. It seems that the use of a continuous porosity variation to represent transition between fluid and porous regions gives physically reasonable results. We will not comment further sigmoid function's parameters which is not the purpose of this paper. We plan instead to showcase applications of continuous porosity variation in case of topology optimization.

\subsection{Gradient computation with adjoint method}

We are going to solve the optimization problem with gradient-based optimization algorithm. The latter requires computation of the gradient of the cost functional with respect to the design variable $\alpha$. To do this we
rely on a continuous adjoint method (see e.g. \cite{gunzburger2003perspectives}) and 
\PH{we compute the adjoint model for general cost functions such as: 
$$
\mathrm{J}\text{ (}\textbf{u}\text{, } \mathrm{p}\text{, } \alpha\text{)}= \int_{\Omega} \mathrm{J}_{\Omega}\text{ (}\textbf{u}\text{, } \mathrm{p}\text{, } \alpha\text{)} \; \mathrm{d} \Omega  +  \int_{\Gamma} \mathrm{J}_{\Gamma}\text{ (}\textbf{u}\text{, } \mathrm{p}\text{, } \alpha\text{)} \; \mathrm{d} \Gamma.
$$
}
The Lagrangian functional related to our optimization problem is written:
\begin{eqnarray*}
&& \mathrm{L}\;(\textbf{u},\mathrm{p},\alpha,\textbf{u}^{*},\mathrm{p}^{*}) =  \\
&&
+ \int_{\Omega} \mathrm{J}_{\Omega}\;(\textbf{u},\mathrm{p},\alpha) \; \mathrm{d} \Omega + \int_{\Gamma} \mathrm{J}_{\Gamma}\;(\textbf{u},\mathrm{p},\alpha) \; \mathrm{d} \Gamma 
\\
&& 
- \int_{\Omega} \mathrm{p}^{*} \; \nabla \cdot \textbf{u} \; \mathrm{d} \Omega 
- \int_{\Omega} \textbf{u}^{*} \cdot \left[- \nabla \cdot(\dfrac{1}{\varepsilon\;\mathrm{Re}}\;\nabla \textbf{u}) \right] \mathrm{d} \Omega 
\\
&&
- \int_{\Omega} \textbf{u}^{*} \cdot \left[ \nabla \cdot (\dfrac{1}{\varepsilon^2}\;\textbf{u}\;\textbf{u}) +\dfrac{1}{\varepsilon^3}\;\textbf{u}\textbf{u}\cdot \nabla \varepsilon+ \nabla \mathrm{p}  \right] \mathrm{d} \Omega \\
&&
- \int_{\Omega} \textbf{u}^{*} \cdot \left[ \beta \;\dfrac{(1-\varepsilon)^{2}}{\varepsilon^{3}\;\mathrm{Re}}\;\textbf{u} + \beta^{1/2} \; \dfrac{\mathrm{c}_{\mathrm{F}}\;(1-\varepsilon)}{\varepsilon^{3/2}} \;\textbf{u}\cdot |\textbf{u}| \right] \mathrm{d} \Omega 
\end{eqnarray*}
\begin{eqnarray*}
&&
- \int_{\Gamma_{\mathrm{w}}}(\textbf{u}\cdot \Phi_{\mathrm{w}})\;\mathrm{d}\Gamma_{\mathrm{w}}\\
&&
- \int_{\Gamma_{\mathrm{in}}}((\textbf{u}-\textbf{u}_{\mathrm{in}})\cdot \Phi_{\mathrm{in}})\;\mathrm{d}\Gamma_{\mathrm{in}}\\
&&
- \int_{\Gamma_{\mathrm{out}}}(\partial_{\textbf{n}}\textbf{u}\cdot \Phi_{\mathrm{out}}+\mathrm{p}\;\varphi)\;\mathrm{d}\Gamma_{\mathrm{out}},
\end{eqnarray*}
where we introduced Lagrange multiplier ($\textbf{u}^*, \mathrm{p}^*, \Phi_{\mathrm{w}}, \Phi_{\mathrm{in}}, \Phi_{\mathrm{out}},\varphi$).  
\PH{It is worth noting that the last three adjoint variables will not contribute to the adjoint model (see e.g. \cite[p. 9, Remark 1]{Ramalingom2018}) and are introduced only to enforce the boundary condtions.}
Denoting by: 
$$
\frac{\partial F}{\partial X}[\delta X]=\lim_{h\to 0}\frac{F(X+h\delta X)-F(X)}{h},
$$
the derivative of a given function $F:X\in \textbf{E}\mapsto F(X)\in\textbf{F}$ for two normed spaces $\textbf{E},\textbf{F}$, the adjoint model can be defined by:
$$
\frac{\partial\mathrm{L}}{\partial (\textbf{u},\mathrm{p})}[\delta\textbf{u},\delta \mathrm{p}]=0,\ \forall \delta\textbf{u},\delta \mathrm{p}.
$$
Some computations then gives:
\begin{eqnarray*}
&& \dfrac{\partial \mathrm{L}}{\partial (\textbf{u},\;\mathrm{p})} \; \left[ \delta \textbf{u},\; \delta \mathrm{p} \right] = \int_{\Omega} \dfrac{\partial \mathrm{J}_{\Omega}}{\partial (\textbf{u},\;\mathrm{p})} \; \left[ \delta \textbf{u},\;\delta \mathrm{p}\right] \; \mathrm{d} \Omega 
\\
&& +\; \int_{\Gamma} \dfrac{\partial \mathrm{J}_{\Gamma}}{\partial (\textbf{u},\;\mathrm{p})} \; \left[ \delta \textbf{u},\;\delta \mathrm{p}\right] \; \mathrm{d} \Gamma +\; \int_{\Omega} \delta \mathrm{p} \; \nabla \cdot \textbf{u}^{*} \; \mathrm{d} \Omega \;
\\
&&
-\; \int_{\Gamma} \delta \mathrm{p} \; \textbf{u}^{*} \cdot \textbf{n} \; \mathrm{d} \Gamma  \\
&&
+\;\int_{\Omega} \delta \textbf{u} \cdot \left(\nabla \mathrm{p}^{*} + \left[\nabla (\textbf{u}^*)\right]^{\mathrm{T}}\dfrac{\textbf{u}}{\varepsilon^2} \right)\; \mathrm{d} \Omega 
\end{eqnarray*}
\begin{eqnarray*}
&&
+\;\int_{\Omega} \delta \textbf{u} \cdot \left( \dfrac{1}{\varepsilon^2} \; (\textbf{u}\cdot \nabla)\textbf{u}^{*}  -\dfrac{1}{\varepsilon^3}\;\textbf{u}^*\;\textbf{u} \cdot \nabla \varepsilon \right)\; \mathrm{d} \Omega 
\\
&&
+\;\int_{\Omega} \delta \textbf{u} \cdot \left( \left[\nabla (\textbf{u})\right]^{\mathrm{T}}\dfrac{\textbf{u}^*}{\varepsilon^2} + 
 \nabla \cdot (\dfrac{1}{\varepsilon\;\mathrm{Re}}\;\nabla \textbf{u}^{*}) \right)\; \mathrm{d} \Omega
 \\
 &&
+\;\int_{\Omega} \delta \textbf{u} \cdot \left(- \left[\nabla (\varepsilon\;\textbf{u})\right]^{\mathrm{T}}\dfrac{\textbf{u}^*}{\varepsilon^3} \right)\; \mathrm{d} \Omega\\
 &&
+\;\int_{\Omega} \delta \textbf{u} \cdot \left( - \beta \;\dfrac{(1-\varepsilon)^{2}}{\varepsilon^{3}\;\mathrm{Re}}\;\textbf{u}^{*}  \right)\; \mathrm{d} \Omega
\\
 &&
+\;\int_{\Omega} \delta \textbf{u} \cdot \left(  - \beta^{1/2} \; \dfrac{\mathrm{c}_{\mathrm{F}}\;(1-\varepsilon)}{\varepsilon^{3/2}} \; \left[|\textbf{u}|\;\textbf{u}^{*}+\dfrac{(\textbf{u}\cdot \textbf{u}^{*})}{|\textbf{u}|}\;\textbf{u}\right] \right)\; \mathrm{d} \Omega
\\
&&
-\;\int_{\Gamma} \delta \textbf{u} \cdot \left(\mathrm{p}^{*}\;  \textbf{n} + (\dfrac{1}{\varepsilon^2}\;\textbf{u}^{*}\cdot \textbf{u})\; \textbf{n} + \dfrac{1}{\varepsilon^2} \left[(\textbf{u}\cdot \textbf{n})\;\textbf{u}^{*}\right] \right) \mathrm{d}\Gamma\\
&&
-\;\int_{\Gamma} \delta \textbf{u} \cdot \left( \dfrac{1}{\varepsilon\;\mathrm{Re}} \nabla \textbf{u}^{*} \;\textbf{n}\right) \mathrm{d}\Gamma\\
&&
+\;\int_{\Gamma}\textbf{u}^{*}\cdot \dfrac{1}{\varepsilon\;\mathrm{Re}}\;\nabla \delta \textbf{u}\;\textbf{n}\;\mathrm{d}\Gamma - \int_{\Gamma_{\mathrm{w}}}(\delta \textbf{u}\cdot \Phi_{\mathrm{w}})\;\mathrm{d}\Gamma_{\mathrm{w}}
\\
&&
- \int_{\Gamma_{\mathrm{in}}}(\delta \textbf{u} \cdot \Phi_{\mathrm{in}})\;\mathrm{d}\Gamma_{\mathrm{in}}\\
&&
- \int_{\Gamma_{\mathrm{out}}}(\partial_{\textbf{n}}\delta \textbf{u}\cdot \Phi_{\mathrm{out}}+\delta \mathrm{p}\;\varphi)\;\mathrm{d}\Gamma_{\mathrm{out}}.
\end{eqnarray*}
and we obtain the following final form of the adjoint problem:
\begin{equation}
\label{eq:adjoint_Omega}
\begin{array}{lll}
- \dfrac{\partial \mathrm{J}_{\Omega}}{\partial \mathrm{p}} = \nabla \cdot \textbf{u}^{*}\hspace{1cm}\text{in}\;\Omega\\[5mm]
- \dfrac{\partial \mathrm{J}_{\Omega}}{\partial \textbf{u}} = \nabla \mathrm{p}^{*} + \left[\nabla (\textbf{u}^*)\right]^{\mathrm{T}}\dfrac{\textbf{u}}{\varepsilon^2} + \dfrac{1}{\varepsilon^2} \; (\textbf{u}\cdot \nabla)\textbf{u}^{*}  -\dfrac{1}{\varepsilon^3}\;\textbf{u}^*\;\textbf{u} \cdot \nabla \varepsilon \\[5mm]- \left[\nabla (\varepsilon\;\textbf{u})\right]^{\mathrm{T}}\dfrac{\textbf{u}^*}{\varepsilon^3} + \left[\nabla (\textbf{u})\right]^{\mathrm{T}}\dfrac{\textbf{u}^*}{\varepsilon^2} + 
 \nabla \cdot (\dfrac{1}{\varepsilon\;\mathrm{Re}}\;\nabla \textbf{u}^{*}) \\[5mm]- \beta \;\dfrac{(1-\varepsilon)^{2}}{\varepsilon^{3}\;\mathrm{Re}}\;\textbf{u}^{*} - \beta^{1/2} \; \dfrac{\mathrm{c}_{\mathrm{F}}\;(1-\varepsilon)}{\varepsilon^{3/2}} \; \left[|\textbf{u}|\;\textbf{u}^{*}+\dfrac{(\textbf{u}\cdot \textbf{u}^{*})}{|\textbf{u}|}\;\textbf{u}\right] \hspace{1cm}\\\text{in}\;\Omega.
\end{array}
\end{equation}
For the adjoint boundary conditions, \PH{we obtain}:


\begin{eqnarray}
\nonumber && \mathrm{on\ }\Gamma_{\mathrm{in}}:\  \textbf{u}^{*} \cdot \textbf{n}-\dfrac{\partial \mathrm{J}_{\Gamma}}{\partial \mathrm{p}} =0 \text{ and } u^{*}_t=0 \text{ \cite{Ramalingom2018}},  \\
&& \mathrm{on\ }\Gamma_{\mathrm{w}}:\ \textbf{u}^*=0, \\
\nonumber && \mathrm{on\ }\Gamma_{\mathrm{out}}:\ \mathrm{p}^{*}\;  \textbf{n} + (\dfrac{1}{\varepsilon^2}\;\textbf{u}^{*}\cdot \textbf{u})\; \textbf{n} + \dfrac{1}{\varepsilon^2} \left[(\textbf{u}\cdot \textbf{n})\;\textbf{u}^{*}\right] \\
\nonumber && \hspace{1.3cm} + \dfrac{1}{\varepsilon\;\mathrm{Re}} \nabla \textbf{u}^{*} \;\textbf{n}-\dfrac{\partial \mathrm{J}_{\Gamma}}{\partial \textbf{u}}=0.
\end{eqnarray}



According to the adjoint method, the gradient of the cost functional $\mathrm{J}\text{(}\textbf{u}\text{, } \mathrm{p}\text{, } \alpha\text{)}$ at some $\alpha$ is given by:
\begin{equation}
\label{eq:gradient_cost_func}
\begin{array}{llll}
-\dfrac{\partial \mathrm{J}_{\Omega}}{\partial \alpha}= - 2\;\dfrac{\partial \varepsilon(\alpha)}{\varepsilon^{3}(\alpha)}\;\textbf{u}\;\textbf{u} : \left[\nabla \textbf{u}^{*}\right]^{\mathrm{T}} \\[5mm]+ \nabla \cdot\left[\dfrac{1}{\varepsilon^3(\alpha)}\;\left[\textbf{u}\; \textbf{u}\right]^{\mathrm{T}}\textbf{u}^{*}\right]   \partial\varepsilon(\alpha) + \dfrac{\partial \varepsilon(\alpha)}{\varepsilon^{2}(\alpha)\;\mathrm{Re}} \;\nabla \textbf{u} : \left[\nabla \textbf{u}^{*}\right]^{\mathrm{T}}
\\[5mm]+ \dfrac{3\;\partial \varepsilon(\alpha)}{\varepsilon^{4}(\alpha)} \; \textbf{u}^{*} \cdot \left(\textbf{u}\; \textbf{u} \cdot \nabla \varepsilon(\alpha)\right) + \textbf{u}^{*} \cdot \textbf{u} \; \dfrac{2\;\beta\; (1-\varepsilon(\alpha))\;\partial \varepsilon(\alpha) }{\varepsilon^{3}(\alpha)\;\mathrm{Re}} \\[5mm]+ \textbf{u}^{*}\cdot \textbf{u}\;\dfrac{3\;\beta\;(1-\varepsilon(\alpha))^{2}\;\partial \varepsilon(\alpha)}{\varepsilon^{4}(\alpha)\;\mathrm{Re}} 
+ \textbf{u}^{*} \cdot \textbf{u}|\textbf{u}|\; \dfrac{\beta^{1/2}\;\partial \varepsilon(\alpha)\;\mathrm{c}_{\mathrm{F}}}{\varepsilon^{3/2}(\alpha)}\\[5mm] + \dfrac{3}{2}\;\textbf{u}^{*} \cdot \textbf{u}|\textbf{u}|\;\dfrac{\beta^{1/2}\;\mathrm{c}_{\mathrm{F}}\;(1-\varepsilon(\alpha))\;\partial \varepsilon(\alpha)}{\varepsilon^{3/2}(\alpha)}
\end{array}
\end{equation}

\begin{equation}
\label{eq:gradient_cost_func_gamma}
\begin{array}{llll}
-\dfrac{\partial \mathrm{J}_{\Gamma}}{\partial \alpha}&=2\;\dfrac{\partial \varepsilon(\alpha)}{\varepsilon^{3}(\alpha)}\;\left[\textbf{u}\;\textbf{u}\right]^{\mathrm{T}} \textbf{u}^{*} \cdot \textbf{n} \\
& - \dfrac{1}{\varepsilon^3(\alpha)}\;\left[\left[\textbf{u}\; \textbf{u}\right]^{\mathrm{T}}\textbf{u}^{*}\cdot \textbf{n}\right] \partial \varepsilon(\alpha)\\
& - \dfrac{\partial \varepsilon(\alpha)}{\varepsilon^{2}(\alpha)\;\mathrm{Re}}\;\left[\nabla \textbf{u}\right]^{\mathrm{T}} \textbf{u}^{*} \cdot \textbf{n}.
\end{array}
\end{equation}
\PH{Note that}  (\ref{eq:gradient_cost_func_gamma}) reduces to: 
$$
\dfrac{\partial \mathrm{J}_{\Gamma}}{\partial \alpha}=0,
$$
if $\partial_\alpha \varepsilon(\alpha)=0$ on boundaries. \PH{This assumption is actually satisfied if we look for optimized porous media either with constant porosity $\varepsilon_0$ or pure fluid on the boundaries of the computational domain.} 


\section{Results and discussion}
\label{sec:results_num}
We illustrate with help of numerical experimentations the contributions of our new macroscopic model in its reduced form (Section \ref{Governing equations}, Eq. (\ref{modelsimplified})) on topology optimization.
Two geometries from literature are investigated:
\begin{itemize}
\item \textit{Geometry 1:} a bend pipe as studied for instance in \cite{Othmer2008,subramaniam2019topology}.
\item \textit{Geometry 2:} a single pipe as studied \PH{for instance in} \cite{borrvall2003topology,Marck2013,Ramalingom2018}.
\end{itemize}
\PH{For both geometric setting,} the computational domain is square-shaped, with an adimensional side $L = 1$. The inlet flow is prescribed with a constant adimensional velocity equal to unity and the Reynolds number is
\PH{fixed to}:
$$
\mathrm{Re} = 1,000.
$$ 
We use the following values for the parameters involved in the regularization function: 
$$
\tau=0.5,\ \alpha_0 = 10\ \mathrm{and}\  \alpha_{max}=20.
$$ 
\PH{We emphasize} the parameter $\tau$ \PH{is} taken small enough to allow more porosity variation which is the purpose of this paper. \PH{Indeed, the larger $\tau$ is, the sharper the interpolation function is (see \cite[section 3]{Ramalingom2018}). Therefore, a smaller value of $\tau$ allows the porosity to take more intermediate values between fluid ($\varepsilon=1$) and porous ($\varepsilon=\varepsilon_0$).}
We propose to solve the topology optimization problem with a steepest descent algorithm where the gradient is computed thanks to the adjoint method introduced in Section \ref{adjoint method}. Here, the design objective is to minimize a power
function, which for the absence of body fluid forces is the dissipated power in the fluid (see e.g. \cite{Borrvall2002}):
$$
    \mathrm{J}_{\Gamma}(\textbf{u},p) = - \int_{\Gamma} (\mathrm{p}+\dfrac{1}{2}\;\textbf{u}^{2})\; \textbf{u} \cdot \textbf{n}\;\mathrm{d}\Gamma.
$$
This cost functional is of interest and frequently used in the literature. In that sense, it is suitable for our study which is centered in the transition between fluid and porous materials. 
\PH{
Note that $\mathrm{J}_{\Omega}=0$. Some computations also yield:
\begin{equation*}
\begin{array}{ll}
-\dfrac{\partial \mathrm{J}_{\Gamma}}{\partial \mathrm{p}}\;\delta \mathrm{p}=\int_{\Gamma} \delta \mathrm{p}\;\textbf{u}\cdot \textbf{n}\;\mathrm{d}\Gamma\\[5mm]
-\dfrac{\partial \mathrm{J}_{\Gamma}}{\partial \textbf{u}}\;\delta \textbf{u}=\int_{\Gamma}(\mathrm{p}+\dfrac{1}{2}\;|\textbf{u}|^{2})\;\delta \textbf{u}\cdot \textbf{n}\;\mathrm{d}\Gamma + \int_{\Gamma} (\textbf{u}\cdot \textbf{n})\;(\textbf{u}\cdot \delta\textbf{u})\;\mathrm{d}\Gamma,
\end{array}
\end{equation*}
and the adjoint boundary conditions are:
\begin{equation}
\label{eq:Adjoint_BC}
\left\{
\begin{array}{ll}
\textbf{u}^*=\textbf{0} \hspace{1cm}\text{on}\;\Gamma_{w}\\[5mm]
u^*_t=0 \text{ and } u^*_n=-u_n \hspace{1cm}\text{on}\;\Gamma_{in}\\[5mm]
-\; \mathrm{p}^{*}\;\textbf{n} - (\dfrac{1}{\varepsilon^2}\;\textbf{u}^{*}\cdot \textbf{u})\;\textbf{n} - (\dfrac{1}{\varepsilon^2}\;\textbf{u}\cdot \textbf{n})\;\textbf{u}^{*} -\dfrac{1}{\varepsilon\;\mathrm{Re}} \nabla \textbf{u}^{*} \;\textbf{n}\\[5mm]= (\mathrm{p}+\dfrac{1}{2}\;|\textbf{u}|^{2})\;\textbf{n} + (\textbf{u}\cdot \textbf{n})\;\textbf{u}\hspace{1cm}\text{on}\;\Gamma_{out}.
\end{array}
\right.
\end{equation}
For the pressure losses cost function, the adjoint model then reduces to Eq. (\ref{eq:adjoint_Omega}) with boundary conditions (\ref{eq:Adjoint_BC}) and the gradient of the cost function can be found in (\ref{eq:gradient_cost_func}),(\ref{eq:gradient_cost_func_gamma}).
}

The main steps of the algorithm consist to compute sensitivities by adjoint method and evaluate the optimality condition. For our simulations we used a SIMPLE algorithm, \PH{to solve both direct and adjoint problems,} and stopped the process when residuals for $\textbf{u} \leq 10^{-5}$ and for $\mathrm{p}\leq 10^{-6}$. The forward problem and the optimization processes are implemented using OpenFOAM. At the beginning of the optimization process the cavity is filling of fluid. The design variables are evaluated by using the conjugated-gradient descent direction method associated to Polack–Ribiere method \cite[p. 5, Figure 2]{Ramalingom2018}. We started the optimization procedure at iteration 1,000 to avoid numerical issues when solving the pressure at the beggining. 
Residuals of the cost functional reached $10^{-7}$ before the end of the SIMPLE algorithm.   


This section begins with topology optimization procedure using an almost impermeable domain which corresponds to a solid of Darcy number equal to $10^{-5}$ \cite{PapoutsisKiachagias2014}. 
Thereafter a comparison when porosity variation is not taken account in topology optimization, as is often the case at present is analysed. This section ends with a study of porosity and particle diameter effects on material distribution.


\subsection{Case of an impermeable domain}

\begin{figure*}
    \centering
    \includegraphics[scale=0.5]{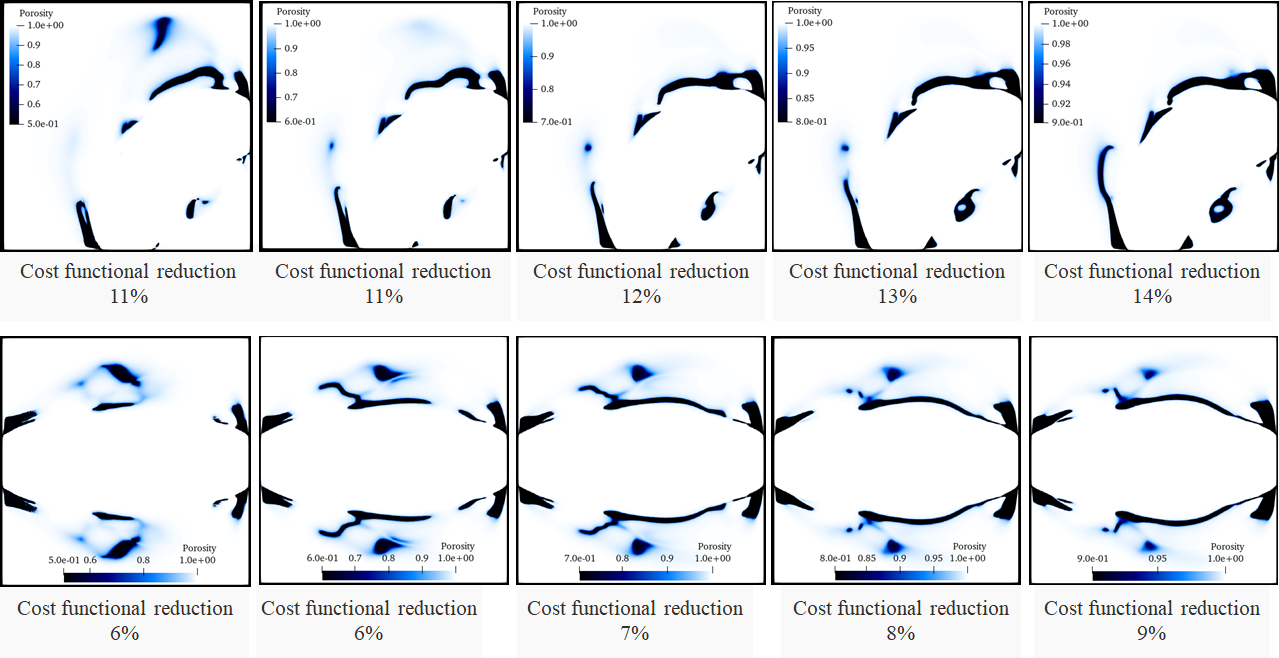}
    \caption{Different designs obtained in case of a constant Darcy number equal to $10^{-5}$ for the Geometry 1 (upper) and Geometry 2 (lower)}
    \label{optim_da_10_5}
\end{figure*}

\begin{figure*}
    \centering
    \includegraphics[scale=0.5]{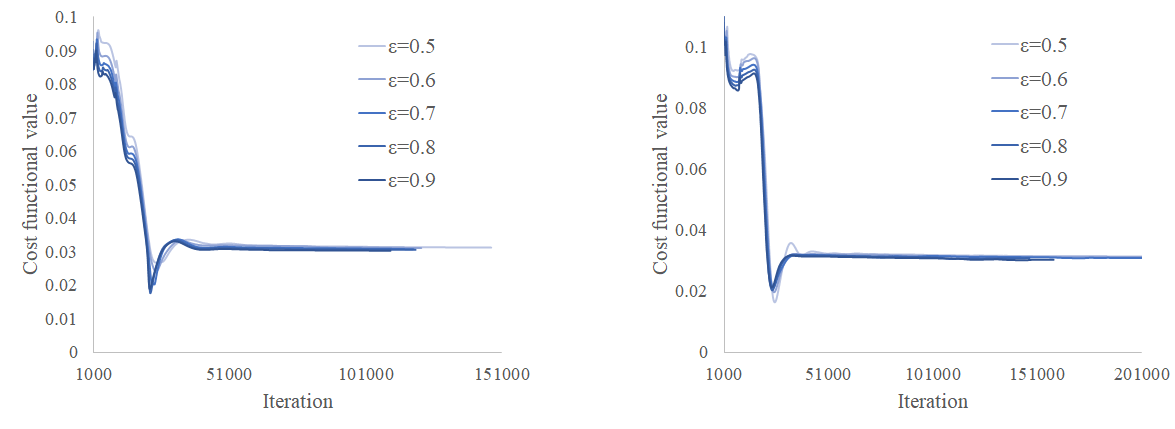}
    \caption{Cost functionals values over iterations in case of constant Darcy number equal to $10^{-5}$ for Geometry 1 (left) and Geometry 2 (right)}
    \label{functionals_da_10_5}
\end{figure*}

Conventionally, a cell who must contain a solid is equivalent to a penalization term sufficiently strong so that porous material added becomes impermeable. Solidified part becomes almost impermeable (see \cite{Olesen2006,PapoutsisKiachagias2014}) for: 
$$
\mathrm{Da} \leq 10^{-5}.
$$
We show in Figure \ref{optim_da_10_5} that with a complete description of a porous medium, considering porosity $\varepsilon$ and particle diameter $\mathrm{d}_\mathrm{p}$ by means of $\beta$ \PH{and} for a fixed Darcy number of $\mathrm{Da}=10^{-5}$, we obtain different \PH{optimized designs}. To that extent, porous material distribution obtained only by specifying permeability is lacking. We provide cost functionals evolutions in Figure \ref{functionals_da_10_5} for the single pipe and the bend pipe. There is a reduction of cost functionals values over iterations until a plateau is reached for each geometrical configuration and for each $(\varepsilon_0,\beta)$ combination. The different combinations are presented in Table \ref{tableau 1} as well as final values of each functionals with and without optimization. Final values are close but lowest ones are obtained for high porosities and high beta values hence small particle diameter values. Final designs obtained after optimization are different. We observe that high values of porosity and beta enhanced added material ratio and provide better cost functionals reduction, up to 9\% for the single pipe and 14\% for the bend pipe. So we can conclude that, topology optimization procedure with Darcy number of $10^{-5}$ results in different final designs depending on the porosity and the particle diameter of the porous medium under study.

\begin{table*}
    \centering
    \caption{Porosity and particle diameter combinations to maintain a constant Darcy number of $10^{-5}$ and final functional value associated}
    \begin{tabular}{c|c|c|c}
     \hline
    $\varepsilon$ & $\beta$ & $\mathrm{J}$ bend pipe (Geometry 1) & $\mathrm{J}$ single pipe (Geometry 2) \\
     \hline
      - & - &  0.0350053 &  0.0333810\\
      $0.9$  &  $\simeq 7.10^6$ & 0.0302537 & 0.0302941\\
      $0.8$  &  $\simeq 1.10^6$ & 0.0305171 & 0.0308692\\
      $0.7$  &  $\simeq 4.10^5$ & 0.0307196 & 0.0309085\\
      $0.6$  &  $\simeq 1.10^5$ & 0.0310652 & 0.0312197\\
      $0.5$  &  $= 5.10^4$ & 0.0311963 & 0.0314046\\
    \end{tabular}
    \label{tableau 1}
\end{table*}

\begin{table*}
    \centering
    \caption{Functional reduction (\%) in taking account porosity variation (a) or not (b).}
    \begin{tabular}{c|c|c|c|c}
     \hline
        $\varepsilon$ & $\beta$ & $\mathrm{Da}$ & Geometry 1 (a) & Geometry 1 (b)\\
         \hline
      $0.3$  &  $1.10^6$ & $6.10^{-8}$ & 18 & 18\\
      $0.3$  &  $1.10^5$ & $6.10^{-7}$ &12 & 12\\
      $0.5$  &  $1.10^6$ & $5.10^{-7}$ &17 & 17 \\
      $0.5$  &  $1.10^5$ & $5.10^{-6}$ &12 & 12 \\
      $0.5$  &  $2.10^4$ & $3.10^{-5}$ &9 & 7  \\
      $0.8$  &  $1.10^6$ & $1.10^{-5}$ &12 & 12 \\
      $0.8$  &  $1.10^5$ & $1.10^{-4}$ &5 & 7  \\
      $0.8$  &  $2.10^4$ & $6.10^{-4}$ &2 & 6  \\
      \hline
    $\varepsilon$ & $\beta$ & $\mathrm{Da}$ & Geometry 2 (a) & Geometry 2 (b) \\
    \hline
      $0.3$  &  $1.10^6$ & $6.10^{-8}$ & 13 & 14\\
      $0.5$  &  $1.10^6$ & $5.10^{-7}$ & 12 & 13\\
      $0.5$  &  $1.10^5$ & $5.10^{-6}$ & 8 & 6\\
      $0.5$  &  $2.10^4$ & $3.10^{-5}$ & 3 & 3\\
      $0.8$  &  $1.10^6$ & $1.10^{-5}$ & 7 & 6\\
      $0.8$  &  $1.10^5$ & $1.10^{-4}$ & 2 & 2\\
      $0.8$  &  $2.10^4$ & $6.10^{-4}$ & 0 & 2\\
    \end{tabular}
    \label{tableau 2}
\end{table*}

\subsection{With and without gradient of porosity}

In this study we have developed a macroscopic flow model in which porosity can spatially vary. Hence porosity coefficients are found within gradient operators. Moreover, terms of gradient of porosity appear. For a convective flow, general model becomes a reduced one because some terms are negligible (Section \ref{model reduction}). Reduced model is able to reproduce same results as the general one in the case presented in Section \ref{porous plug flow}. We have made numerical experimentations without taking into account porosity variation. This resulted in our case by the absence of the term $(\dfrac{1}{\varepsilon^3}\;\textbf{u} \textbf{u} \cdot \nabla \varepsilon)$ in the reduced form of the momentum equation. Cost functional’s improvement percentages are presented in Table \ref{tableau 2} for the single pipe and the bend pipe in comparison with results obtained in taking account porosity variation. 

Several values of $\varepsilon$ and $\beta$ were tested to have broad range of comparison. Percentages of reduction of objective functionals are very similar, with or without gradient of porosity. Maximal deviation is $4\%$. For the bend pipe, the value of $\varepsilon=0.5$ / $\beta=2.10^4$ allows an improvement of $2\%$ in favour of the model with gradient of porosity. In contrast, for $(\varepsilon,\beta)=(0.8,1.10^5)$ and $(\varepsilon,\beta)=(0.8,2.10^4)$ the model without considering gradient of porosity allows for better improvement (respectively $7\%$ and $6\%$) instead of respectively ($5\%$ and $2\%$). 

For the single pipe, taking account gradient of porosity is attractive regarding objective functional’s values for $(\varepsilon,\beta)=(0.5,1.10^5)$. Otherwise, for other values, deviations are of the order of $1\%$. It may be noted that $(\varepsilon,\beta)=(0.8,2.10^4)$ does not allow any enhancement for the objective functional in the case of single pipe in taking account porosity variation. The latter observation is of importance since optimization’s algorithm have \PH{still} added materials inside the computational domain. 

\begin{sidewaysfigure*}
    \centering
    \includegraphics[scale=0.7]{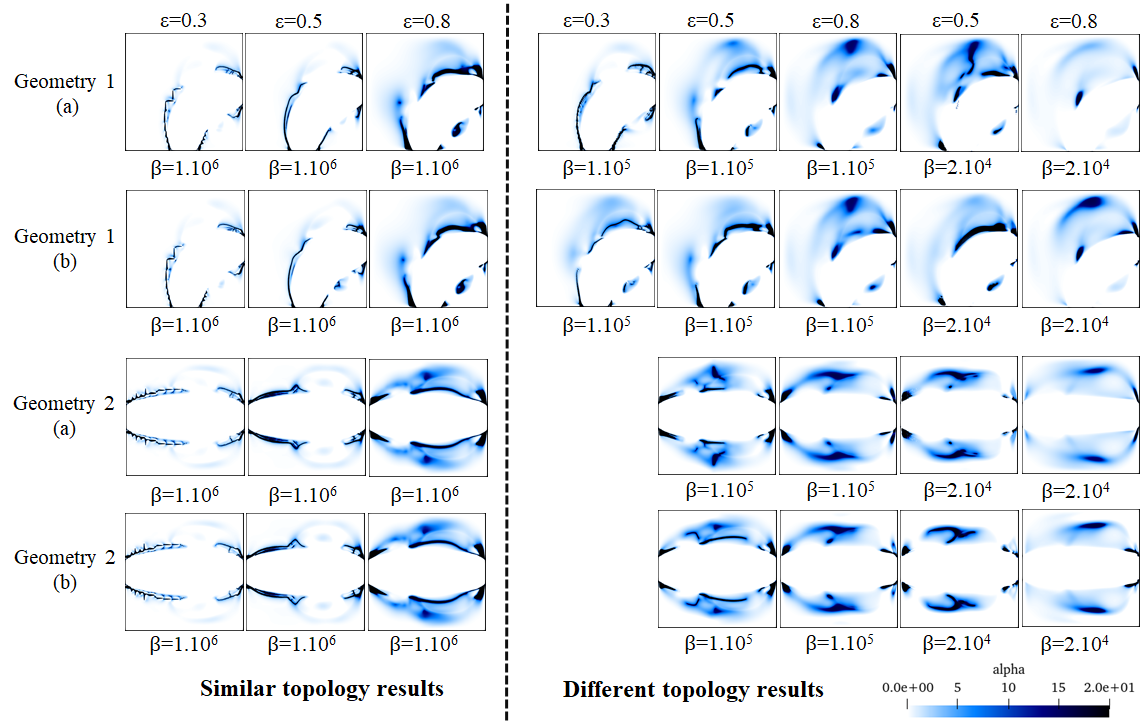}
    \caption{Distribution of the design variable $\alpha$. In taking account porosity variation (a) or not (b). $\alpha \in \left[0,\; \alpha_{max}=20 \right]$ and $\alpha=0$ correspond to fluid.}
    \label{All results}
\end{sidewaysfigure*}

Distributions of the variable design $\alpha$ are presented in Figure \ref{All results}. For $\alpha=0$ (white colour) we have fluid and for $\alpha=20$ (black colour) we have porous materials. In terms of designs obtained, the parameter $\beta$ and thus the particle diameter is the key indicator if there is a need in considering porosity variation. For $\beta=1.10^6$ designs obtained are similar. However, for all other configurations, designs obtained after optimization are different.

\subsection{Effect of porosity value}

Numerical experiments for $\varepsilon = 0.3-0.5-0.8$ were carried out. For $\beta=1.10^6$ cost functionals present reduction of 18 and 12\% respectively for $\varepsilon=0.3-0.8$ in the case of the bend pipe (Table \ref{tableau 2}). It should be noted that percentages of reduction are identical with or without taking account porosity variation. In the case of single pipe, we have an enhancement of $13\%$ and $7\%$ with gradient of porosity and $14\%$ and $6\%$ without gradient of porosity. When porosity increases, objective functionals reduction declines. Up to half less in the case of single pipe when porosity varies from $0.3$ to $0.8$. Nevertheless, when porosity varies from $0.3$ to $0.5$ and $\beta=1.10^6$, cost functionals present only variation of $1\%$ for the bend pipe and the single pipe. This trend is confirmed by $\beta=1.10^5$ where in the case of the bend pipe, changing porosity from $0.3$ to $0.5$ has no effect on the cost functionals amelioration. For $\beta=2.10^4$, it is interesting to note that without taking account porosity variation, \PH{reduction} of cost functionals are imperceptible (near $1\%$) when porosity varies from $0.5$ to $0.8$. But the model who takes account porosity variation presents a significant gap from $9\%$ to $2\%$ for the bend pipe and $3\%$ to $0\%$ for the single pipe. 

\PH{Regarding} the designs obtained, we observe that the larger the porosity is, the higher is the field of actualisation of the variable of conception. This can be due to an accentuation of zones which must be penalized \PH{resulting to more porous} materials added by the algorithm. Consequently, the optimization’s algorithm has more difficulty to penalize key areas. It follows that cost functional reduction is less attractive.

\begin{figure}
    \centering
    \includegraphics[scale=0.5]{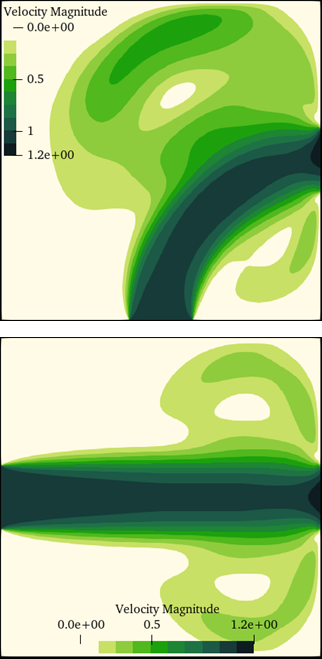}
    \caption{Velocity magnitude without optimization for the bend pipe (Geometry 1) and for the single pipe (Geometry 2).}
    \label{velocity_sans_optim}
\end{figure}

\subsection{Effect of particle diameter value}

Numerical experiments for $\beta = 1.10^6-1.10^5-2.10^4$ were finally carried out. For $\varepsilon=0.5$ we observe a decrease of cost functionals reduction once $\beta$ decline. Percentages improvement of cost functionals are $17-12\%$ and $9\%$ for the bend pipe and $12-8\%$ and $3\%$ for the single pipe when taking account porosity variation. For $\varepsilon=0.8$ we make the same observation, percentages are $12-5\%$ and $2\%$ for the bend pipe and $7-2\%$ and $0\%$ for the single pipe when taking account porosity variation. This is logical because, for a given stream tube section, a straight flow is naturally less mechanical energy consuming than a tortuous flow path.

Regarding designs obtained, they are totally different \PH{from those for a constant porosity} according to the value of $\beta$. It can be concluded that $\beta$ and thus particle diameter has a great impact on the final design after optimization. This finding is reinforced by the fact that a high $\beta$ coefficient always corresponds to a better reduction of the objective functional. In the case of the bend pipe, for $\varepsilon=0.3$, when $\beta$ \PH{goes} from $1.10^6$ to $1.10^5$, percentages of reduction vary from $18\%$ to $12\%$ while for $\varepsilon=0.5$, percentages of amelioration vary from $17\%$ to $12\%$. In that case, it is clear that porosity has no influence while $\beta$ is indeed the important parameter.

\begin{sidewaysfigure*}
    \centering
    \includegraphics[scale=0.65]{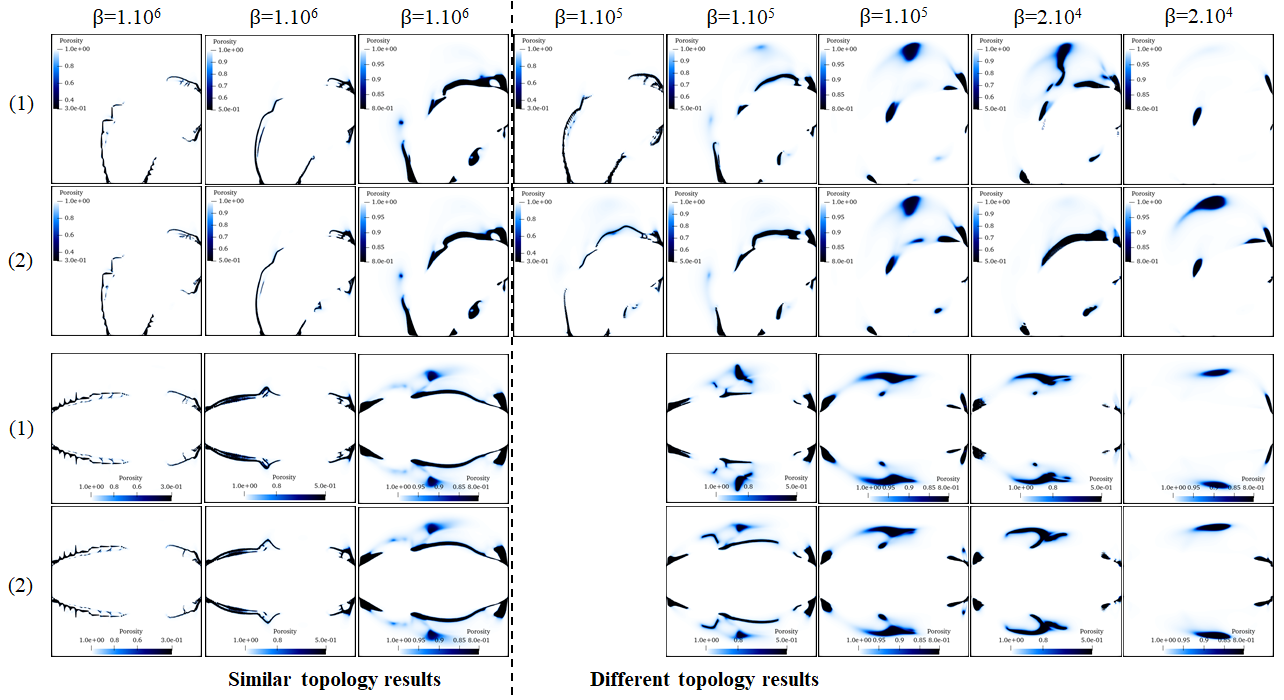}
    \caption{Porosity distribution.}
    \label{porosity distribution}
\end{sidewaysfigure*}

\begin{sidewaysfigure*}
    \centering
    \includegraphics[scale=0.7]{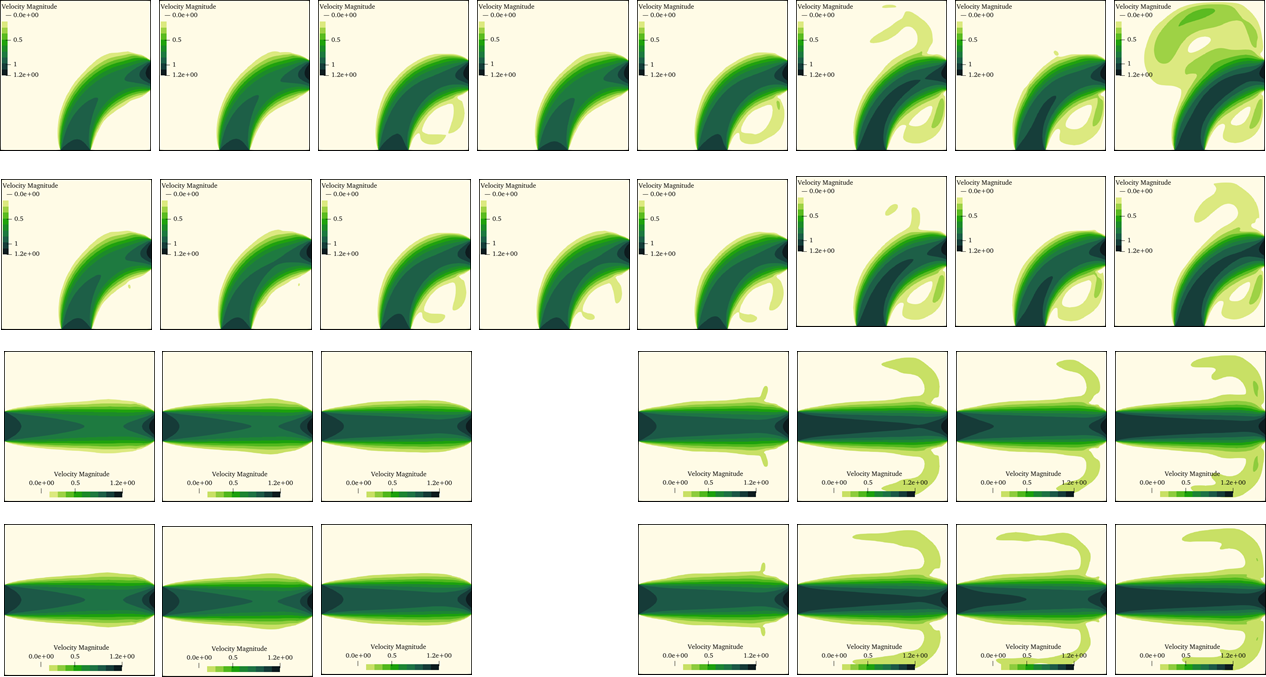}
    \caption{Velocity distribution.}
    \label{velocity distribution}
\end{sidewaysfigure*}

\AB{

\textbf{Remark:}
\textit{
As seen from Figure \ref{velocity_sans_optim}, there are recirculations generated by the main flow between the two openings. These secondary flows are driven by shear. The two vortex structures, once established, maintain the main flow in its position.
However, these secondary flows, although they are useful to the main flow to maintain themselves, cost the fluid system energy. It seems therefore natural that the optimization algorithm adds material (see Figure \ref{porosity distribution}) to inhibit these inefficient areas while preserving the overall shape of the main flow (see Figure \ref{velocity distribution}).}
}

\section{Conclusions and future works}

\PH{
In this paper, we have modelled Newtonian incompressible flow through porous media having spatially varying porosity. Thanks to dimensional analysis, we obtained a reduced, hence simpler, mathematical model that takes into account spatial variations of porosity. We then justified both our numerical code and our model by comparing our results 
on a benchmark from the litterature, namely a porous insert inside a fluid channel.
}
We have shown that the method outlined here can correctly predict flows with or without a porous matrix. We have also shown that it is possible to optimize a flow by minimizing the mechanical power loss using a porous medium in which the distribution of $\mathrm{d}_{\mathrm{p}}$ and $\mathrm{\textbf{k}}$ are heterogeneous. 
\PH{It is worth noting that the adjoint model has been derived for a general cost function  and can then be applied without any changes to other cost functions.}

\PH{Regarding the perspectives, we emphasize that the mathematical model and the general adjoint method developed in this paper} make it possible to consider the use in future of real porous media to optimize fluidic systems. 
\PH{In addition, since this paper developed modelling of flow through variable media, it could be very interesting to consider now topology optimization problems with so-called multi-materials 
\cite{zuo2017multi,hvejsel2011material}, hence optimized porous media with piecewise-constant porosity. A last very interesting (yet difficult) perspective may be to extend the results obtained this paper to time-dependent flows in spatially/time varying porous media.   
}

\bibliographystyle{unsrt}  
\bibliography{references}
\end{document}